\documentclass[conference]{IEEEtran}
\usepackage{cite}
\usepackage{amsmath,amssymb,amsfonts}
\usepackage{algorithmic}
\usepackage{graphicx}
\usepackage{textcomp}
\usepackage{float}
\usepackage{mathtools}

\usepackage{graphicx}
\graphicspath{ {./} }
\usepackage{xcolor}

\begin{document}

\title{Measurement Based Evaluation and Mitigation of Flood Attacks on a LAN Test-Bed\thanks{This research has been supported by the European Commission H2020 Program through the IoTAC Research and Innovation Action, under Grant Agreement No. 952684.}} 
	
\author{\IEEEauthorblockN{Mohammed Nasereddin and Mert Nak\i p}
	\IEEEauthorblockA{Institute of Theoretical \& Applied Informatics\\
		Polish Acad. Sci., 44-100 Gliwice, Poland\\
	\{mnasareddin,mnakip\}@iitis.pl}
	\and
	\IEEEauthorblockN{Erol Gelenbe}
	\IEEEauthorblockA{Institute of Theoretical \& Applied Informatics\\
		Polish Acad. Sci., 44-100 Gliwice, Poland\\
	\& Ya\c{s}ar University, Bornova, Turkey,\\
	\& Lab. I3S Universit\'{e} C\^{o}te d'Azur, Nice, France}
seg@iitis.pl}
	

\maketitle

\begin{abstract}
The IoT is vulnerable to network attacks, and Intrusion Detection Systems (IDS) can provide high attack detection accuracy and are easily installed in IoT Servers. However, IDS are seldom evaluated in operational conditions which are seriously impaired by attack overload. Thus a Local Area Network test-bed is used to evaluate the impact of UDP Flood Attacks on an IoT Server, whose first line of defence is an accurate IDS. We show that attacks overload  the multi-core  Server and paralyze its IDS. Thus a mitigation scheme that detects attacks rapidly, and drops packets within milli-seconds after the attack begins, is proposed and experimentally evaluated.
\end{abstract} 

\begin{IEEEkeywords}
	Internet of Things, Local Area Networks, Cybersecurity, Random Neural Networks, G-Networks, UDP Flood Attacks, Intrusion Detection and Mitigation
\end{IEEEkeywords}

\section{Introduction}
 Denial of service (DoS) disables systems or networks by flooding them with huge streams of requests,  causing reputational damage, with financial and productivity losses \cite{ISCIS2018}. In the last year, a $150\%$ increase in such attacks occurred worldwide \cite{Staff_2023}, targeting the IoT,  industrial control systems, power grids and transportation systems \cite{Oke,Spilios,rajesh2020detecting,al2021ddos}, with DoS and Botnet attacks spreading  via their victims, who then become attackers \cite{Statt,Tushir_impactsOfMirai,Sinanovic}. Flood attacks \cite{Cloudflare2} overwhelm networks with large numbers of forged-source address packets, causing delayed or lost data, inaccurate or incomplete readings and overload   \cite{mirkovic2008accurately}. Thus these threats require effective Intrusion Detection Systems (IDS).

While much of the literature on IDS evaluates them under ideal conditions where attack traffic is treated as data, this paper compares ``ideal''  results about attack detection (AD) algorithms, with system performance measurements in a LAN environment.  Section \ref{Related} reviews related work, Section \ref{ExperimentalSetup} describes the experimental test-bed that we use and Section \ref{AADRNN} briefly discusses the IDS and its performance under ideal and real-world conditions. 
Section \ref{Experiments} presents measurements during different UDP Flood attacks, and summarizes improvements resulting from simple attack mitigation. Finally, Section \ref{Conclusions} concludes the paper and outlines future work directions.

\subsection{Related Work} \label{Related}

Using test-beds to evaluate IDSs 
was recommended in early work \cite{mirkovic2009test}, and
several  test-beds for cyber-physical systems, industrial control and IoT environments have been described \cite{waraga2020design}, such as real-time test-bed for cyber-physical systems \cite{vellaithurai2015development}, power systems  \cite{ashok2016testbed}, and wind farms \cite{singh2020testbed}.   In \cite{kaouk2018testbed}, a semi-physical test-bed for ICSs was proposed, while in \cite{annor2018development}, a low-cost Smart Grid test-bed for SCIDS systems was evaluated for TCP flood attacks. Test-beds for SCADA systems are discussed in \cite{ghaleb2016scada,tesfahun2016scada,reutimann2022simulating}.
In \cite{park2018test}, a test-bed using six NetFlow tools for collecting, analyzing, and displaying data was proposed for HTTP-GET flood attacks in a WAN. In \cite{arthi2021design}, the impact of attack datasets on IoT systems is discussed and real-time data collection for DNS amplification is investigated, while DoS attacks on software defined networks are discussed in \cite{wright2019testbed}. In \cite{sontakke2022impact}, DoS attacks on an autonomous vehicle test-bed are described.
Attack datasets are reviewed in \cite{park2018test,arthi2021design}, while the present paper uses the MHDDoS repository \cite{MHDDOS} of real-world DoS attacks.

\section{Experimental Setup}\label{ExperimentalSetup}

\begin{figure}[t!]
\centering
\includegraphics[height=4cm,width=7cm]{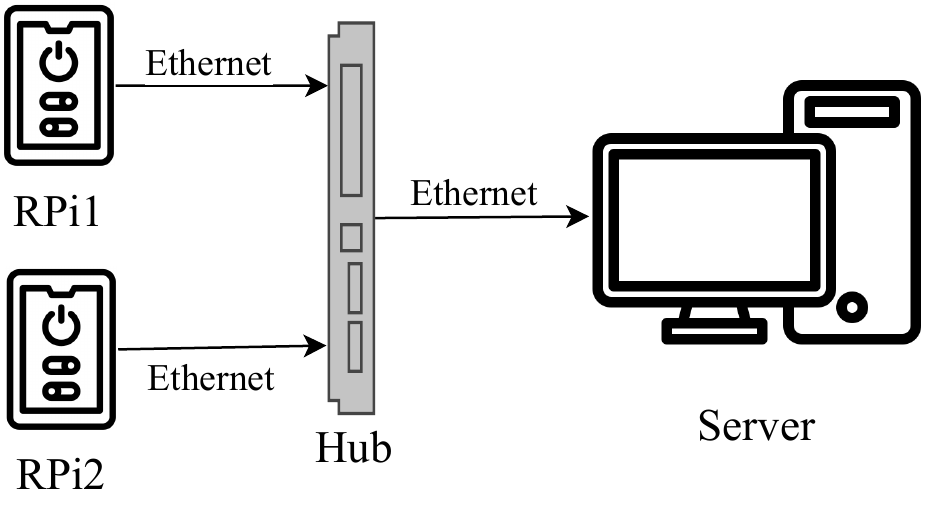}
\caption{Testing Environment using Ethernet for communications, with Raspberry Pi machines acting as forwarders of normal and attack traffic, and an Intel 8-Core Processer used as a  Server to process incoming packet traffic and run the IDS   algorithm.}
\label{Fig1}
\end{figure}

In this paper,  a physical test-bed environment is constructed to evaluate IDS   algorithms in more realistic conditions, with  an arbitrary number of linked devices, multiple sources of normal and attack traffic, and a  Server that supports the UDP Protocol for incoming traffic, runs the IDS   algorithm, and processes the incoming packets' contents. The devices that generate benign or attack traffic, are embodied by  Raspberry Pi $4$ Model B Rev $1.2$ machines (RPi$1$ and RPi$2$), each with a $1.5$GHz ARM Cortex-A$72$ quad-core processor and $2$GB LPDDR$4-3200$ SDRAM, running Raspbian GNU/Linux $11$ (bullseye), a Debian-based operating system for the Raspberry Pi hardware. A  Server with eight Intel Core i$7-8705$G processors acts as the receiver of the packet traffic and is responsible for detecting the attack and storing the arriving packets. It has $16$GB of RAM, a $500$GB hard drive, and runs Linux $5.15.0-60-$ generic $66-$Ubuntu SMP, an Ubuntu-based operating system. Its cores run at $3.10$GHz. UDP traffic is carried over Ethernet connections between all  devices via the Hub in Figure \ref{Fig1}, without ACKs or error recovery \cite{kumar2012survey}.

\subsection{The IDS and its Ideal Performance} \label{AADRNN}

\begin{figure}[h]
	\centering
	\includegraphics[scale=0.75]{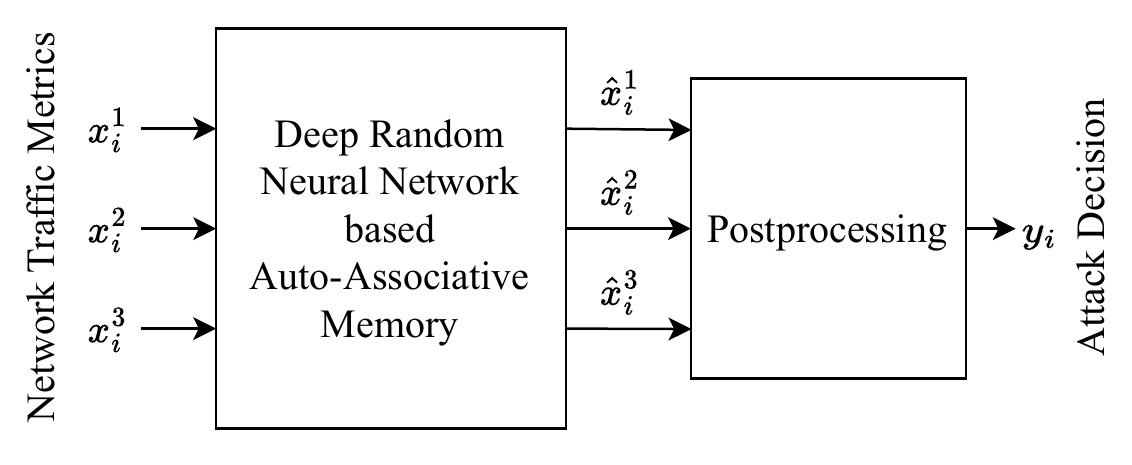}
	\caption{The structure of the IDS  system that computes the decision variable $y_i$ from the network traffic metrics $[x_i^1, x_i^2, x_i^3]$ with the DRNN based Auto-Associative Random Neural Network (AAD  RNN) and the postprocessing module.}
	\label{fig:IDS}
\end{figure}

\begin{figure}[h!]
	\hspace{-0.4cm}\includegraphics[width=9.5cm,height=4cm]{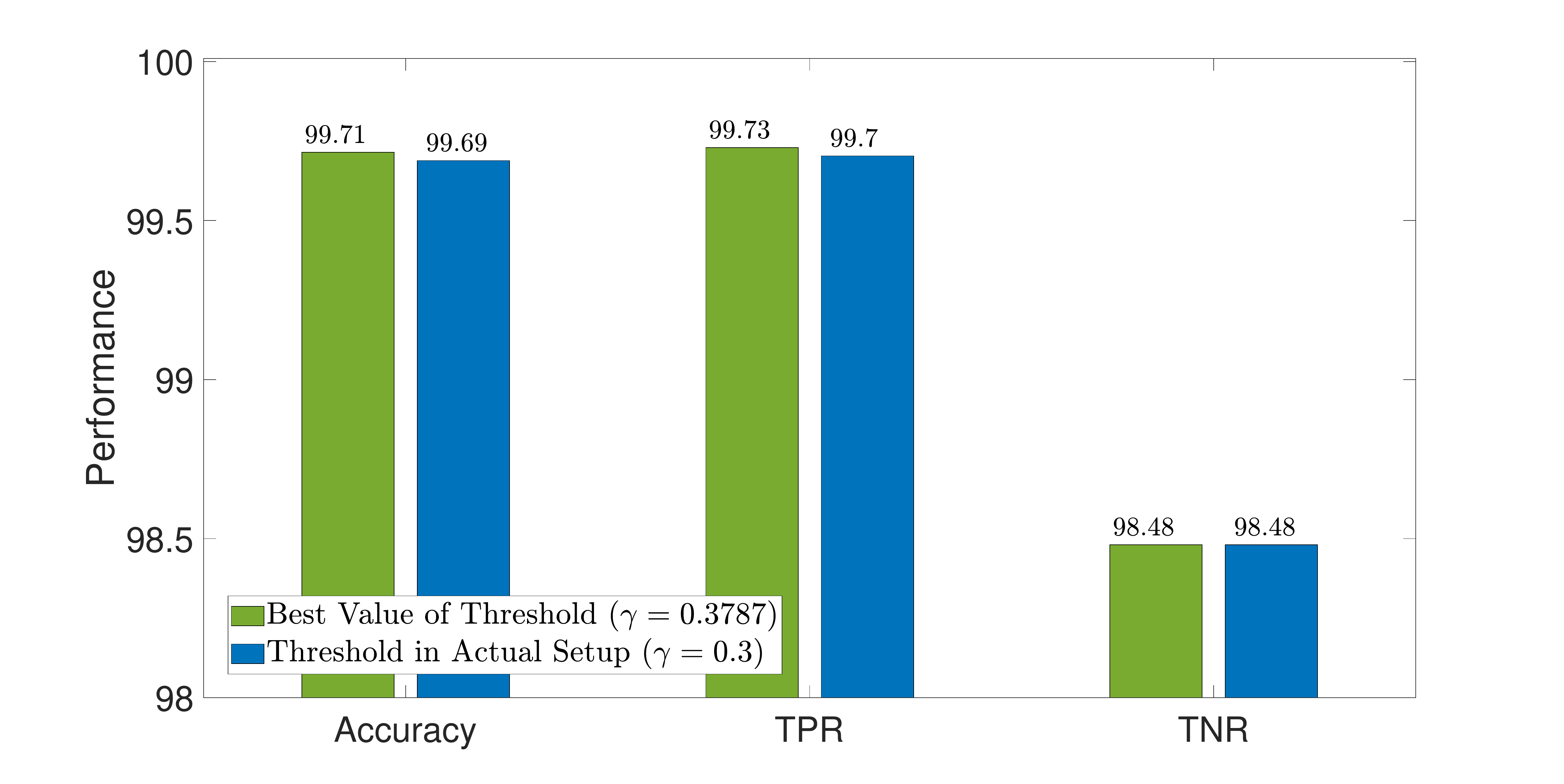}
	\caption{The performance of the IDS  with $\gamma = 0.3$, and compared with the best value of $\gamma=0.3787$, is evaluated for Accuracy, TPR, and TNR, in an experiment where RPi2 starts a UDP Flood attack lasting $10$ seconds.}
	\label{AADRNN_performance}
\end{figure}

\begin{figure}[h!]
	\hspace{-0.4cm}\includegraphics[width=9.5cm,height=4cm]{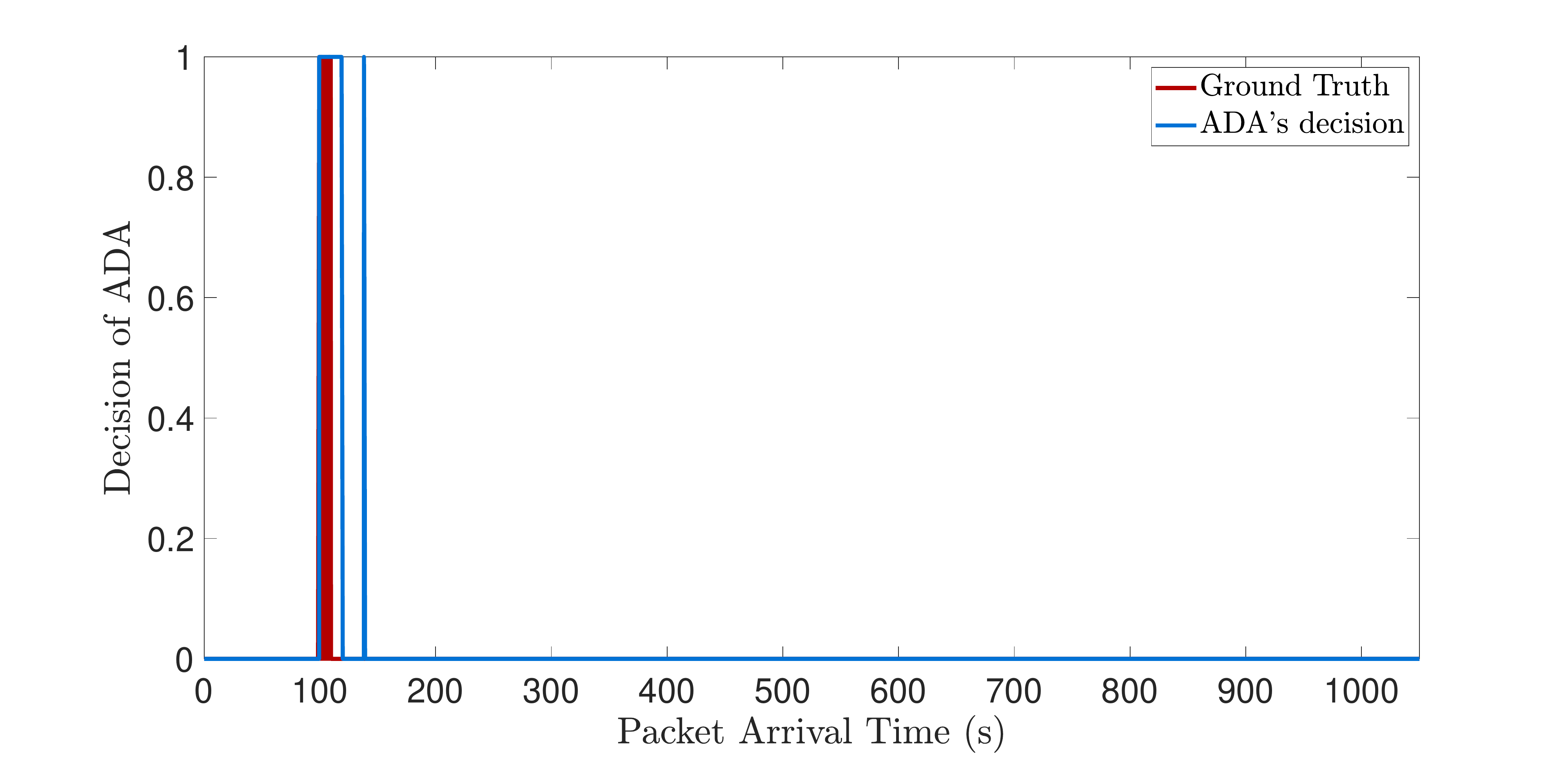}
	\caption{The IDS's binary decisions are shown for $\gamma = 0.3$, when the RPi2 starts a UDP Flood attack lasting $10$ seconds.}
	\label{fig:binary_decision}
\end{figure}

The IDS  used in this paper, based on the Deep Random Neural Network (DRNN) \cite{ISCIS2018-2}, is shown in Figure~\ref{fig:IDS}. It {\bf learns from the } first $500$ benign packets received by the  Server,  with metrics $x^i=[x_i^1, x_i^2, x_i^3]$ related to successive sets of packets. The IDS
predicts the expected metrics: $\hat{x}_i = [\hat{x}_i^1, \hat{x}_i^2, \hat{x}_i^3]$,  and the difference between its input and the prediction yields
the decision variables $y_i$ (attack or non-attack), which are also used to update the algorithm's weights.  The IDS uses the DRNN \cite{deep2016}, a Random Neural Network \cite{RNN1} with soma-to-soma triggering between neurons. This IDS provides accurate detection with different datasets \cite{mirai2021, incremental2022, compromised2022,simultaneous2022}, with the excellent statistical performance shown in Figure~\ref{AADRNN_performance}, which reports the Accuracy, TPR, and TNR, for a $10$ second attack. This IDS attains high accuracy both when a fixed threshold $\gamma=0.3$ is used, and for the best threshold $\gamma=0.3787$, exhibiting  $99.7 \%$ Accuracy and TPR, while TNR is $98.48 \%$. These results are not significantly different when the attack lasts for $60$ seconds. Note that Figure~\ref{fig:binary_decision} shows that, due to the decision delay, the IDS  may raise an alarm just after an attack ceases.

\section{Experiments with Normal and Attack Traffic} \label{Experiments}

\begin{figure}[h!]
\centering
\includegraphics[height=4cm,width=8cm]{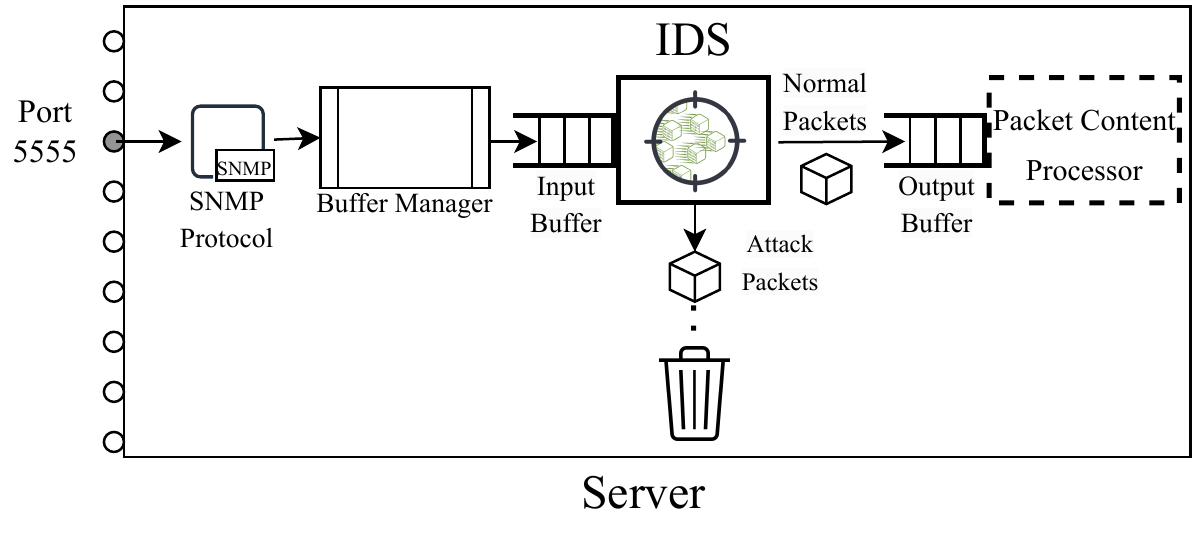}
\caption{Schematic organization of the  Server that supports the IDS.  Mitigation is based on triggering ``packet drop'' decisions for all packets in  the IDS   Input Buffer, when it detects a majority of attack packets among the most recent $20$ packets. The  IDS then resumes testing the incoming packets, and the decision and mitigatiion process is repeated.}
\label{Fig2}
\end{figure}

In Figure \ref{Fig2}, the  Server receives packets from linked devices on port $5555$, which are then passed to the buffer manager by the network protocol, and queued for analysis at the IDS, whose decisions are based on the average of a batch of $10$ successive packets, that  are being classified as normal or attack traffic.  Packets classified as ``normal'' are forwarded to the packet content processor for the rest of the Server's operations. 

When there is no attack, each RPi device  generates normal IP packet traffic containing the device's own CPU temperature and transmits it to the  Server every ($1$) second using UDP. RPi$2$ is programmed to send both normal and attack traffic via random sampling, and generates attacks
from the public repository MHDDoS \cite{MHDDOS}: each $1$ second, it initiates a UDP Flood attack with a probability of $0.10$, or sends normal traffic packet with a probability of $0.90$. RPi$1$ only sends normal traffic.

%

\begin{figure}[h!]
\centering
   \includegraphics[height=3cm,width=9cm]{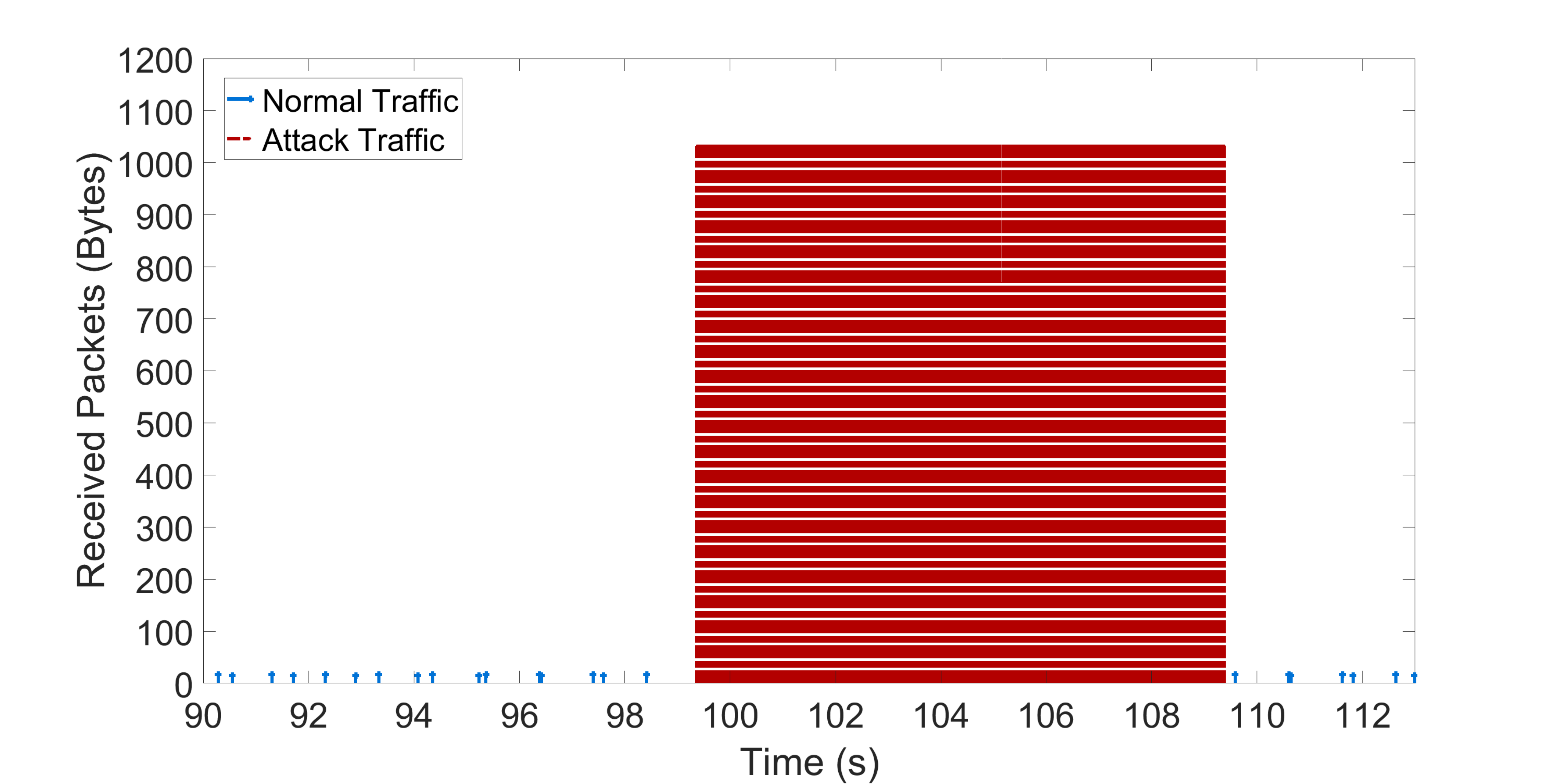}
    \caption{The difference between the normal and attack traffic on the  Server that is targeted by a UDP Flood attack.}
    \label{fig:NormalvsAttack}
\end{figure}
\begin{figure}[h!]
    \centering
   \includegraphics[height=4cm,width=9cm]{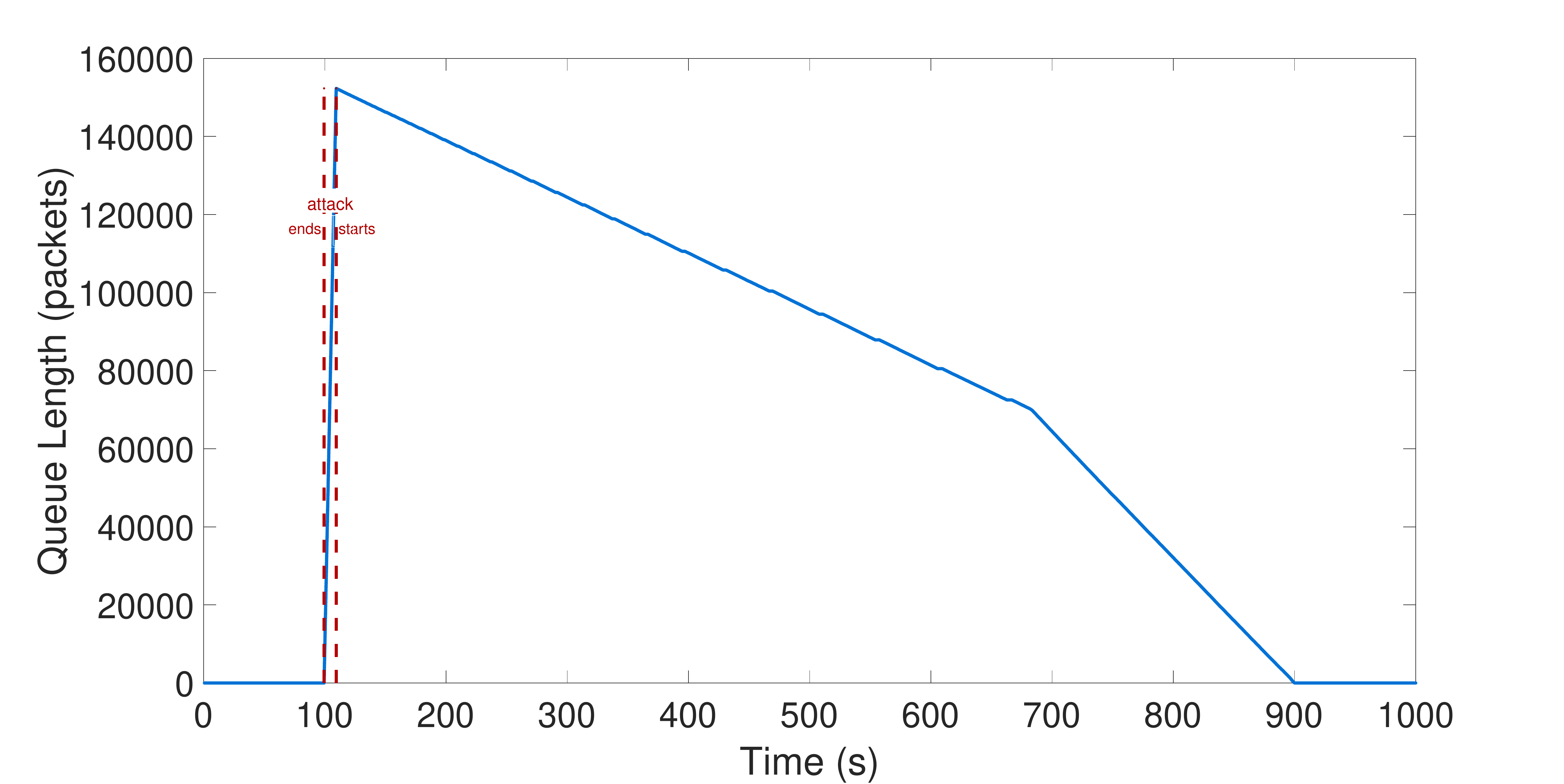}\\
\centering
   \includegraphics[height=3cm,width=9cm]{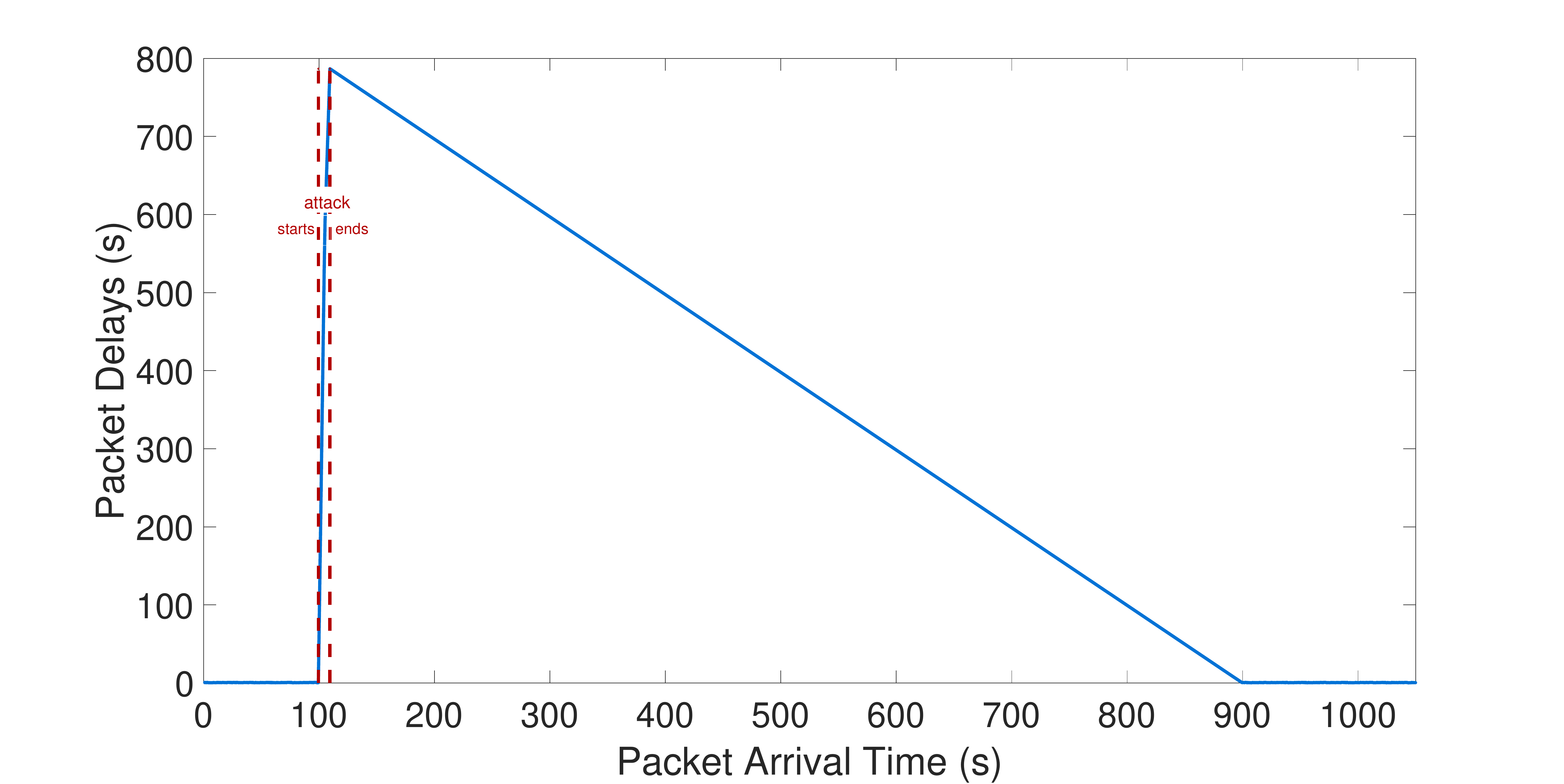}
 \caption{The top figure shows the queue length infront of the IDS  in a $10$ second attack, and the vertical red dashed lines show the active duration of the attack originating in the compromised device RPi$2$. The bottom figure shows the delay before the packet is processed by the IDS.}
	\label{fig:queue_length}
\end{figure}
\begin{figure}[h!]
\centering
   \includegraphics[height=4cm,width=9cm]{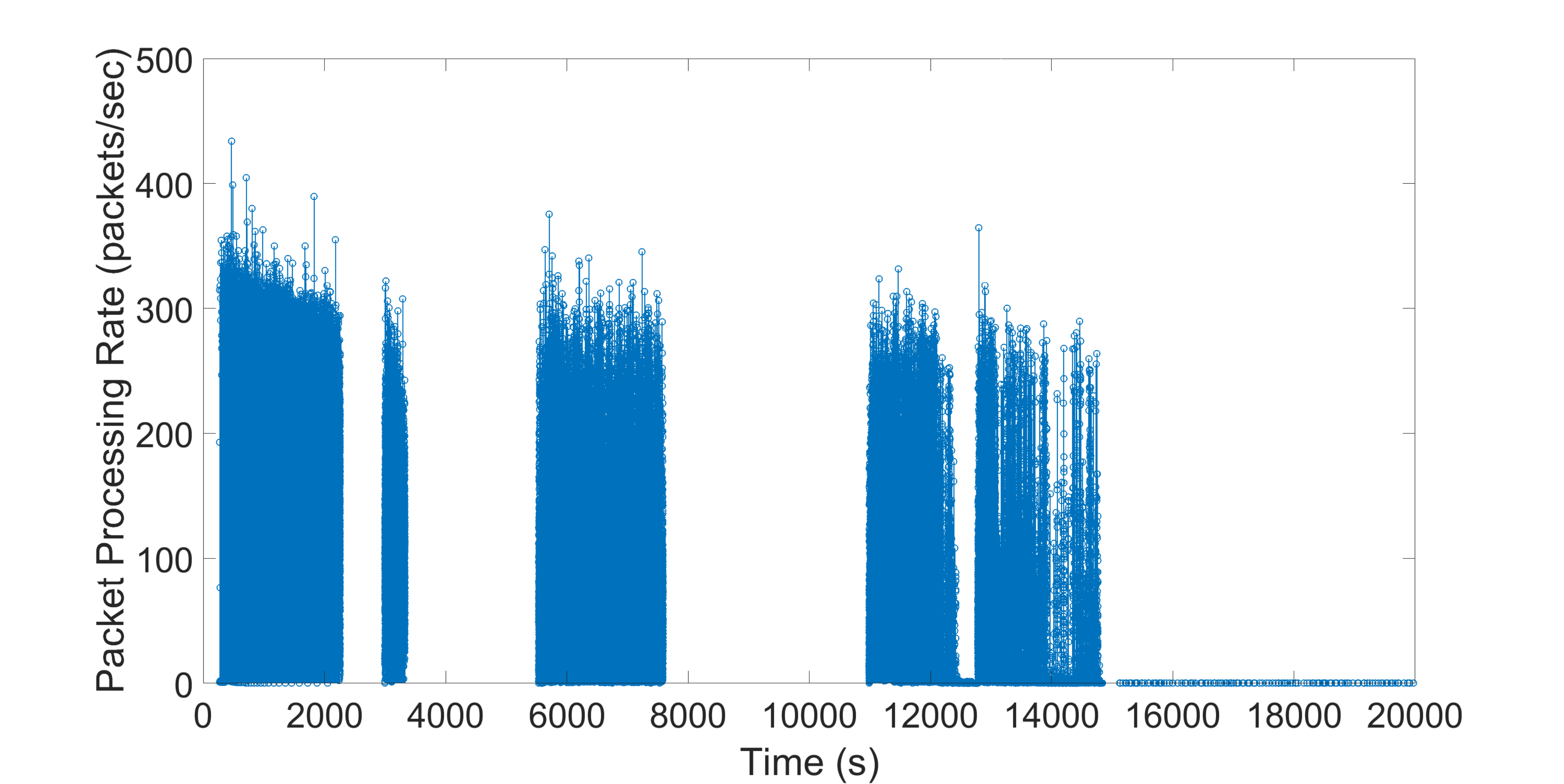}\\
\centering
    \includegraphics[height=4cm,width=9cm]{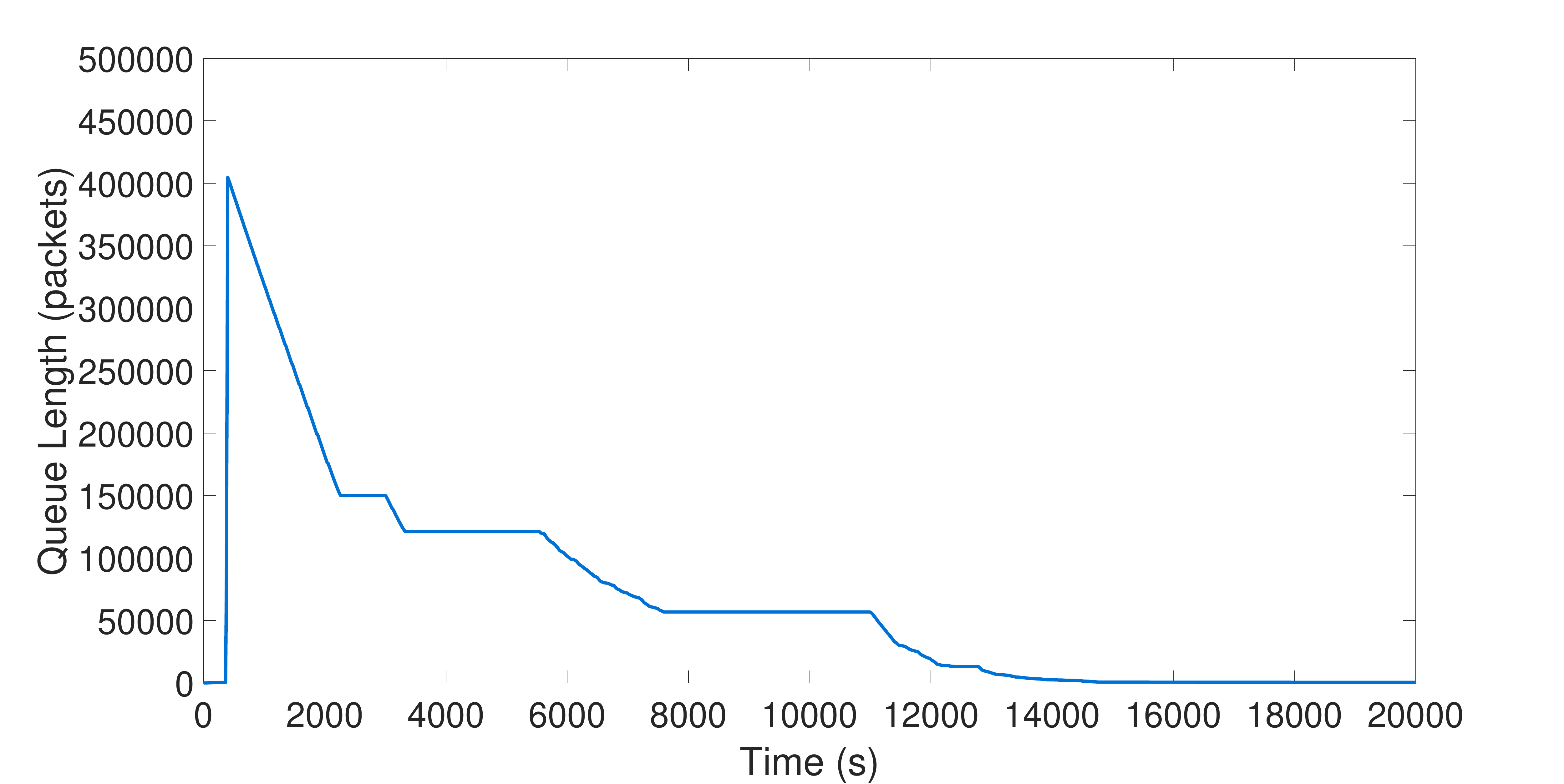}
    \caption{At the top, the effect of a $60$ second UDP Flood attack on the IDS  traffic processing rate in packets per second, is shown when the attack duration is $60$ seconds. The corresponding packet queue length infront of the IDS  is shown at the bottom.}
    \label{fig:DoS1}
\end{figure}

Figure \ref{fig:NormalvsAttack}, displays an intense flow of $1032$ byte attack packets during $10$ seconds, while ``normal'' traffic typically consists of two small packets sent each second, for each IoT device. Figure \ref{fig:queue_length} shows that the packet queue length at the  Server rises sharply infront of the IDS, with a gradual decrease after the attack. Figure \ref{fig:DoS1} shows (Above) the effect of a $60$ second attack on the Server's intermittently paralyzed packet processing rate ($y$-axis in packets/sec), and (Below) the resulting huge input queue length.  Figures \ref{fig:queue_length} and \ref{fig:DoS1}  show that a $10$ second attack floods the packet queue, and the IDS  completes the analysis of the accumulated packets over a long $15$ minute period, while for a $60$ second attack, the IDS is intermittently paralyzed and its analysis can last $5.85$ hours, since the Server's  cores are  busy handling the incoming traffic. Thus the effect of an attack typically lasts far longer than the attacker's activity.

\begin{figure}[h!]
	\centering\includegraphics[height=3.5cm,width=9cm]{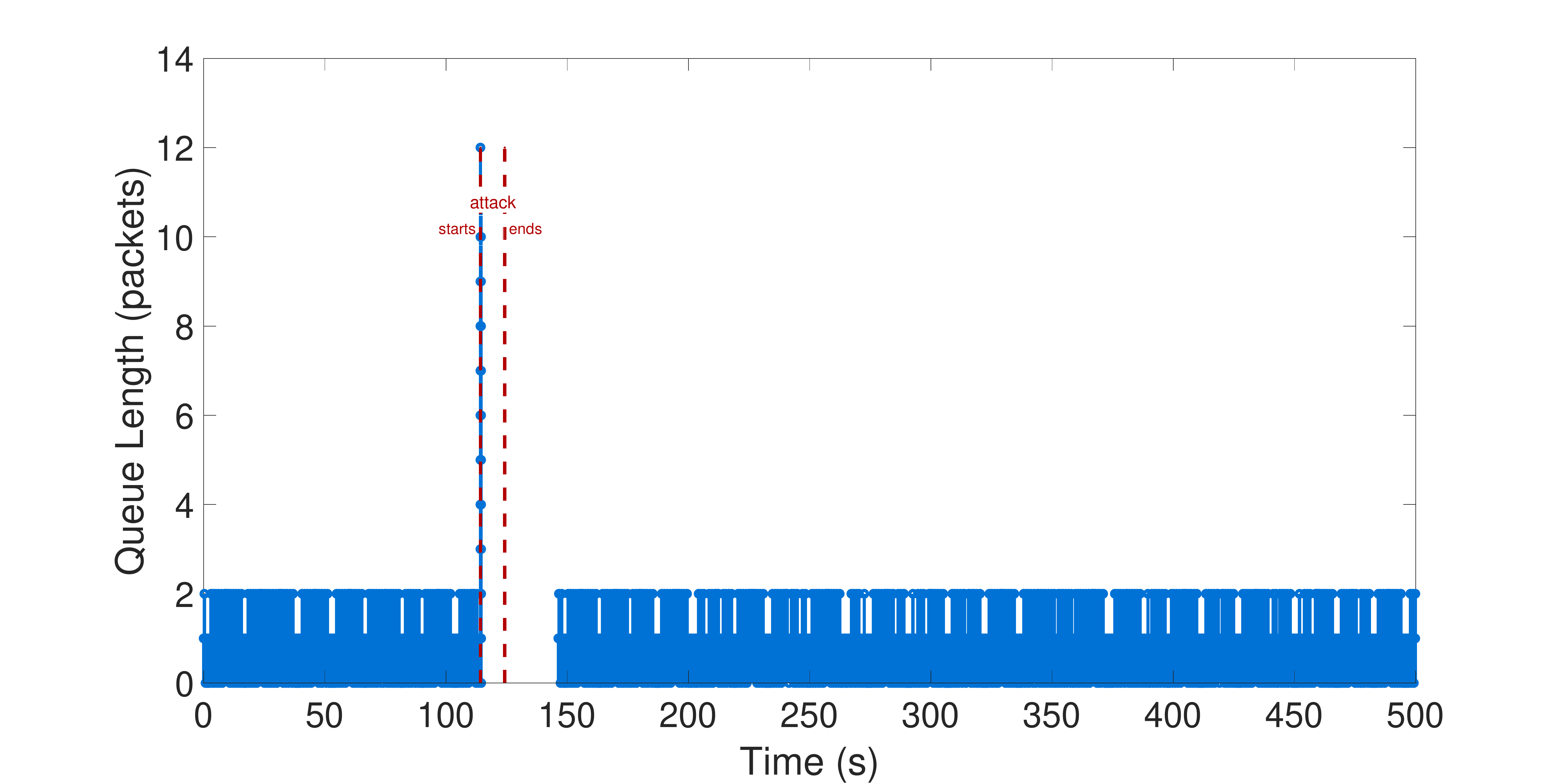}
	\caption{During the $10$ second attack, the decision to drop packets results in a very short packet queue length, avoiding  Server and IDS  paralysis.}
	\label{fig:mitigation_10sec}
\end{figure}

We now report system measurements when a novel mitigation action occurs: if the IDS  decides that the majority of the $20$ most recent packets are attack packets, then the input buffer contents and all incoming packets within the next $30$ seconds are dropped. This action is repeated at the end of the $30$ second window. Figure~\ref{fig:mitigation_10sec} displays the queue length in the input buffer when the attack mitigation is performed during a UDP Flood attack which lasts $10$ seconds: the queue length increases until the IDS  processes $20$ packets and decides to empty the buffer, and we observe that the attack is mitigated successfully.



\begin{figure}[h!]
	\centering\includegraphics[height=3.5cm,width=9cm]{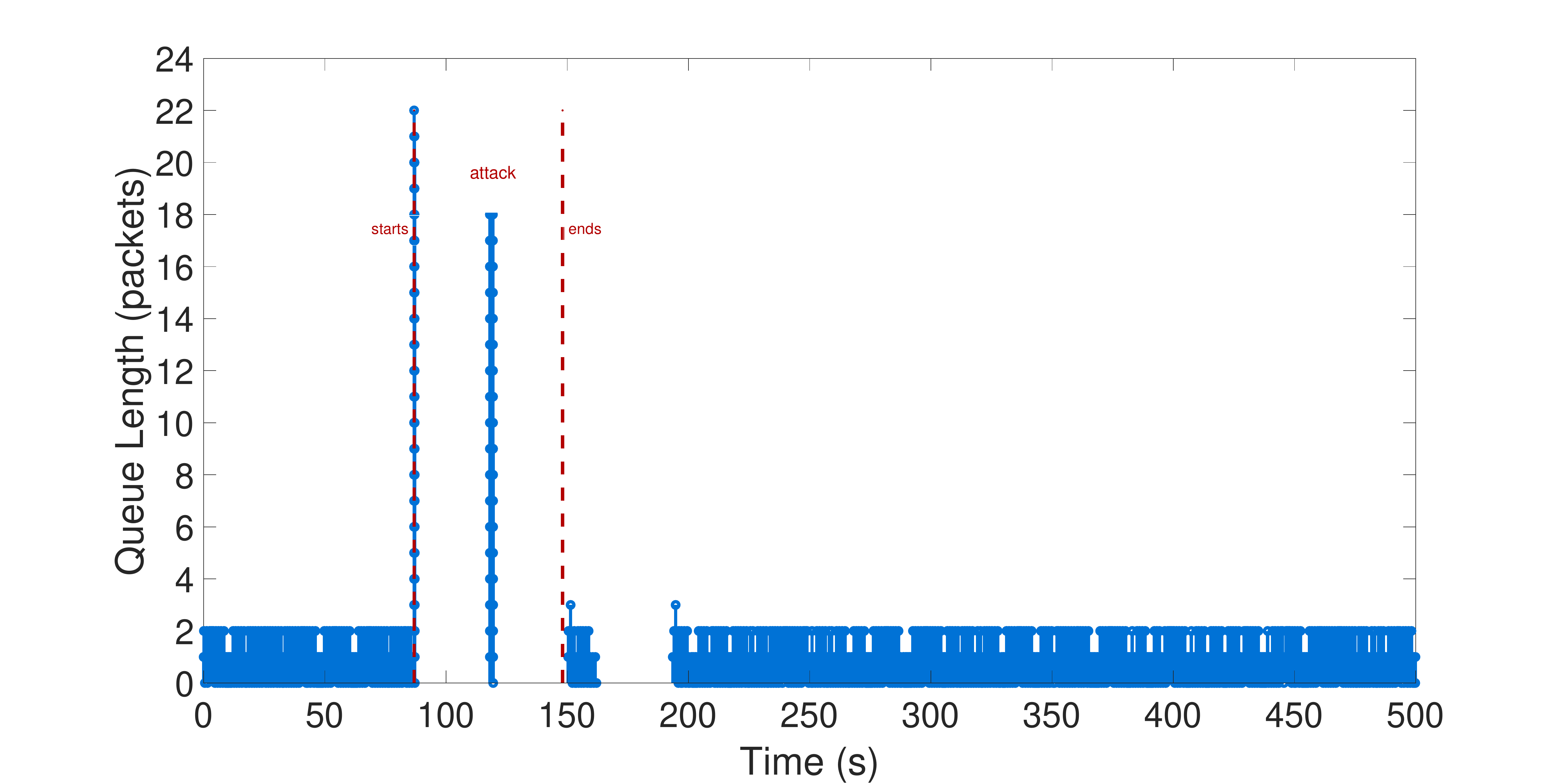}
	\caption{The figure shows that during the attack's $60$ seconds, the mitigation decision occurs twice, with the second mitigation occurring after a detection that takes place between $162$ and $192$ seconds.}
	\label{fig:mitigation_60sec}
\end{figure}

Figure~\ref{fig:mitigation_60sec} shows the queue length when an attack that lasts $60$ seconds is mitigated: the buffer length increases to $22$ packets, which is small compared to the value without mitigation shown in Figure~\ref{fig:DoS1}: the mitigation decision was taken twice, the second time between $162$ to $192$ seconds after the start of the experiment, and the Server could then operate normally without being paralyzed.

\section{Conclusions and Future Work} \label{Conclusions}

Despite the high accuracy of an IDS installed on a  Server that receives traffic from devices in a LAN network test-bed subjected to UDP Flood attacks, we observe that while short attacks are accurately detected,  longer attacks may be detected with greater delay due to Server overload.  
Thus fast mitigation is proposed to discard attacking traffic with rapidly taken decisions. 
Future work will study mitigation to optimize the traffic that is discarded, the frequency with which the IDS analyzes incoming traffic, as well as the minimization of benign traffic loss, and the system's energy consumption \cite{Pernici}.

\medskip
\noindent {\bf Acknowledgment} Support from the European Commission's H2020 IoTAC Project GA No. 952684, and H2020 DOSS Project GA No. 101120270, is gratefully acknowledged.

\bibliographystyle{ieeetran}
\bibliography{mybib,references_arnn_conference}

\end{document}